\documentclass{emulateapj}

\usepackage{amsmath}
\usepackage{apjfonts}
\usepackage{amssymb}
\usepackage{natbib}
\usepackage{multirow}
\usepackage{mathrsfs}
\usepackage{color}
\shorttitle{Surface Detonations in DD Systems Triggered by Accretion Stream Instabilities}
\shortauthors{Guillochon, Dan, Ramirez-Ruiz \& Rosswog}

\begin{document}

\title{Surface Detonations in Double Degenerate Binary Systems Triggered by Accretion Stream Instabilities}

\author{James Guillochon\altaffilmark{1}, Marius Dan\altaffilmark{2}, Enrico Ramirez-Ruiz\altaffilmark{1}, and Stephan Rosswog\altaffilmark{2}}

\altaffiltext{1}{Department of Astronomy and Astrophysics, University
  of California, Santa Cruz, CA 95064, USA} \altaffiltext{2}{School of
  Engineering and Science, Jacobs University Bremen, Campus Ring 1,
  28759 Bremen, Germany}

\begin{abstract}
We present three-dimensional simulations on a new mechanism for the detonation of a sub-Chandrasekhar CO white dwarf in a dynamically unstable system where the secondary is either a pure He white dwarf or a He/CO hybrid. For dynamically unstable systems where the accretion stream directly impacts the surface of the primary, the final tens of orbits can have mass accretion rates that range from $10^{-5}$ to $10^{-3} M_{\odot}$ s$^{-1}$, leading to the rapid accumulation of helium on the surface of the primary. After $\sim 10^{-2}$ $M_{\odot}$ of helium has been accreted, the ram pressure of the hot helium torus can deflect the accretion stream such that the stream no longer directly impacts the surface. The velocity difference between the stream and the torus produces shearing which seeds large-scale Kelvin-Helmholtz instabilities along the interface between the two regions. These instabilities eventually grow into dense knots of material that periodically strike the surface of the primary, adiabatically compressing the underlying helium torus. If the temperature of the compressed material is raised above a critical temperature, the timescale for triple-$\alpha$ reactions becomes comparable to the dynamical timescale, leading to the detonation of the primary's helium envelope. This detonation drives shockwaves into the primary which tend to concentrate at one or more focal points within the primary's CO core. If a relatively small amount of mass is raised above a critical temperature and density at these focal points, the CO core may itself be detonated.
\end{abstract}


\section{Introduction}

White dwarfs (WDs), the end point of stellar evolution for most stars, are extremely common, with about $10^{10}$ of them residing within the Milky Way \citep{Napiwotzki:2009p3570}. WDs are frequently observed in binary systems with normal stellar companions, and, less frequently, with compact stellar companions. Double degenerate (DD) systems are those in which the companion is another WD. Typically, DD systems are formed via common envelope evolution \citep{Nelemans:2001p1711}. Often, the final result of this evolutionary process is a binary consisting of a carbon-oxygen (CO) WD and a lower-mass helium WD companion \citep{Napiwotzki:2007p1710}.

For decades, Type Ia supernovae (SNe) have been employed as standard candles. Still, even the preferred mechanism exhibits substantial variability \citep{Timmes:2003p2073, Kasen:2009p3427}. Complicating this issue further is that WDs may not need to exceed the Chandrasekhar limit to be capable of exploding; other mechanisms include collisions between WDs in dense stellar environments such as the cores of globular clusters \citep{Rosswog:2009p3556}, tidal encounters with moderately massive black holes \citep{Rosswog:2008p3059, Rosswog:2009p3553, RamirezRuiz:2009p3071}, or as we are introducing in this paper, the accretion of dense material from a He WD companion.

At the distance of tens of WD radii, gravitational radiation can bring two WDs close enough for the less-massive WD (secondary) to overfill its Roche lobe and begin transferring mass to the more-massive WD (primary). Because of the mass-radius relationship of WDs, mass transfer is often unstable and can eventually lead to a merger \citep{Marsh:2004p1880, Gokhale:2007p2850}. For binary mass ratios close to unity the circularization radius $R_{\rm h}$ drops below the radius of the primary $R_{1}$, and thus the accretion stream will directly impact the primary's surface.

Most studies of dynamically unstable DD systems focus on the evolution of the post-merger object and, assuming that the final object has a mass larger than the Chandrasekhar limit, on how the merged remnant may eventually lead to a Type Ia SNe \citep{Livio:2000p3540, Yoon:2007p1003}. These models all assume that the rapid accretion that precedes coalescence is uneventful. In this \textit{Letter}, we present three-dimensional hydrodynamics simulations that demonstrate that explosive phenomena are an inevitable results of the high accretion rates characteristic of the final stages of a merger in a DD system. In some cases, these explosive phenomena may lead to the complete detonation of the CO primary.

\begin{figure*}[tb]
\centering\includegraphics[width=0.5\linewidth,clip=true,angle=-90]{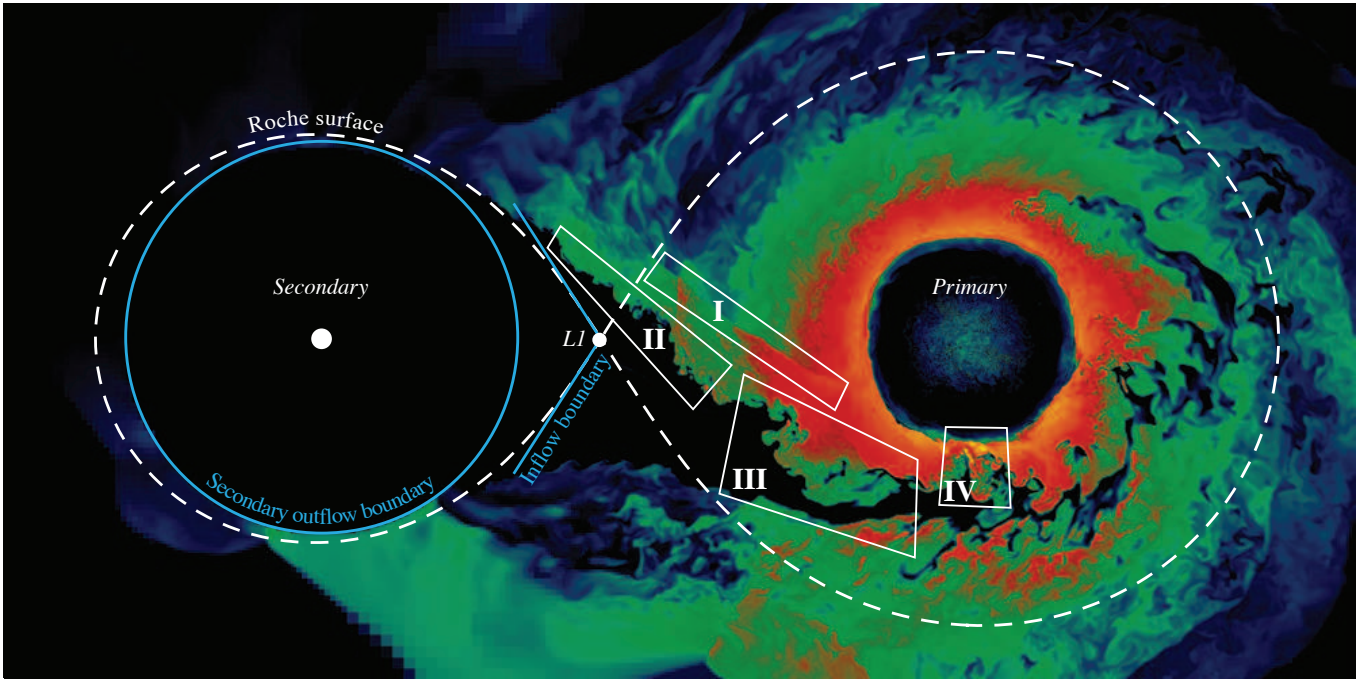}
\caption{Setup of run Ba in FLASH, with several annotated regions of interest. The color scheme shows $\log T$ through a slice of the orbital plane. The cyan circle shows the spherical outflow boundary condition centered about the secondary, while the cyan wedge shows a cross-section of the cone used as the mass inflow boundary. The dashed contour shows the system's Roche surface. The white boxes with labels are described in Section \ref{accinsta}.}
\label{slicefig}
\end{figure*}

\section{Numerical Method and Initial Models}
The lead-up to the merger has only been studied recently due to the sensitivity of the accretion to the transfer of angular momentum. Several groups have simulated these mergers in three dimensions using particle-based codes \citep{Benz:1990p542, Rasio:1995p3148, Segretain:1997p3143, Guerrero:2004p3317, Dan:2008p1716, Dan:2009p3560}, with the latter being able to resolve the final $\sim100$ orbits of the binary prior to coalescence. SPH simulations, with their excellent angular momentum conservation, are well-suited to determine the binary orbital evolution. However, because the SPH particles have a fixed mass and typical accretion rates are $\gtrsim 10^{-5}$ the total mass of the system per orbit, the accretion stream is only resolved by hundreds of particles at best. Consequently, SPH approaches are unable to resolve any of the accretion stream's detailed structure, including how the stream impacts the surface of the primary.

Eulerian codes are able to provide high resolution based on any combination of local and global simulation properties, and thus are ideal for systems where the regions of interest do not overlap with the regions of highest density \citep{New:1997p451, Swesty:2000p452}. Unfortunately, it is difficult to accurately simulate the overall orbital evolution for many orbits in grid-based codes due to the non-conservation of angular momentum \citep{Krumholz:2004p2236}, although a careful choice of coordinate system has led to recent success \citep{DSouza:2006p1559, Motl:2007p3574}. Because the stability of the system depends on both the structure of both the primary and the secondary, the grid-based simulations that investigate stability must fully resolve both stars, and thus the stream and the primary are not very well-resolved in these simulations.

Here we follow a hybrid approach that combines the strengths of both SPH- and grid-based methods. We follow the impact of the mass transfer onto the orbital evolution by performing an SPH simulation (Dan et al. 2009, in prep.) and use the obtained, time-dependent results for orbital separation and $\dot{M}$ as boundary conditions for FLASH \citep{Fryxell:2000p440}. This combination allows us to realistically determine the dynamic stability of the binary while not sacrificing our spatial resolution of the accretion itself.

In FLASH, the primary is built under the assumption of spherical hydrostatic equilibrium and a constant temperature of $5 \times 10^{5}$ K, and is mapped into the simulation explicitly. The secondary is {\it not} mapped directly into the grid and is taken to be a point mass surrounded by a spherical outflow boundary condition. The simulation is performed in a non-inertial frame rotating with angular frequency $\omega = \sqrt{G(M_{1}+M_{2})/a^{3}}$ about the barycenter. The primary's gravity is calculated using a multipole expansion of its potential, and because the quadrupolar contribution of the secondary is negligible, we approximate the secondary's gravitational field as a monopole. Because the choice of frame and the presence of the secondary changes the hydrostatic configuration of the primary, we first allow the primary to slowly relax in the presence of the secondary's potential by removing a fraction of its kinetic energy every time-step over many dynamical timescales.

The nuclear composition of the fluid is evolved via the 13 element $\alpha$-chain network of \cite{Timmes:1999p3513}. Pressures and temperatures are calculated using the Helmholtz equation of state \citep{Timmes:2000p447}. The accretion rate provided by the SPH calculation is used to generate a quasi-hydrostatic inflow boundary condition to replicate the flow of matter from the secondary to the primary (Figure \ref{slicefig}). The boundary we use is cone-shaped with the apex located at {\it L1}, and mass is added to the cone under the assumption that the stream is in hydrostatic equilibrium in the direction perpendicular to the barycentric line \citep{Lubow:1976p3408}. The flow is directed towards the barycenter of the system at velocity $v = c_{\rm s} \sim 10^{8}$ cm/s, the sound speed within the secondary. In all simulations presented here, the stream is assumed to be pure He. The four simulations that we have performed are summarized in Table \ref{outcomes}.

\begin{table*}[tb]
\caption{Simulation results.}
\centering
\leavevmode
\begin{minipage}{\textwidth}
\centering
\begin{tabular}{|l|l|l|l|l|l|l|l|l|l|l|l|}
\hline
Run & $t_{\rm evol}$\tablenotemark{1} & $l_{\min}$\tablenotemark{2} & \multicolumn{2}{l|}{Primary} & \multicolumn{2}{l|}{Secondary} & $M_{\rm torus,\max}$\tablenotemark{3} & Surf. & Mass ejected& \multicolumn{2}{l|}{Burning yield ($10^{-3} M_{\odot}$)}\\ \cline{4-7}\cline{11-12}
&(s)&($10^{6}$ cm)&$M$ ($M_{\odot}$)&Type&$M$ ($M_{\odot}$)&Type&($10^{-2} M_{\odot}$) &det.&($10^{-2} M_{\odot}$)& $20 \le A \le 32$ & $A \ge 36$\\
\hline
A &1885& 4.9& 0.67  & CO & 0.45 & He & $14$ & No & - & - & -\\
B &434& 3.7& 0.9 & CO & 0.45 & He & $9.7$ & Yes & $5.9$ &$11$ & $1.9$\\
Ba\tablenotemark{4} &25& 1.8 & 0.9 & CO & 0.45 & He & $11$ &	No & - & - & -\\
C &335& 3.4 & 0.9 & CO & 0.6 & He/CO & $5.0$ & Yes & $6.0$& $16$ & $5.6$\\
\hline
\end{tabular}
\label{outcomes}
\tablenotetext{1}{Total evolution time.}
\tablenotetext{2}{Minimum grid cell size.}
\tablenotetext{3}{Maximum mass of the helium torus accumulated during the run; this is either the value just before detonation or at the end of the run if there was no detonation.}
\tablenotetext{4}{Run Ba is initialized from a checkpoint produced by run B at $t = 403 s$.}
\end{minipage}
\end{table*}

\section{Accretion Stream Instabilities}\label{accinsta}

In a binary system undergoing mass transfer, the fate of the accretion stream is determined by ratio $R_{1}/R_{\rm h}$. For all three of the binaries we describe $R_{\rm h} < R_{1}$ and thus the stream would directly impact the surface of the primary assuming the stream was collisionless. However, the stream can eventually be deflected by the ram pressure of the thick torus of helium that gradually accumulates on the primary's surface, preventing a direct impact.

As the sound speed within the torus is far smaller than the Keplerian rotation velocity, the collision between the torus and the stream is highly supersonic ($M > 10$), which leads to the development of a standing shock (Figure \ref{slicefig}, region I). This shock establishes pressure equilibrium across the stream-torus interface. Because of the temperature increase of the material falling from {\it L1} to the surface of the primary, the torus consists of material heated to a temperature close to the virial temperature of $GM_{1}m_{\rm p}/R_{1}k_{\rm b} \sim 10^{8}$ K. Radiation pressure becomes competitive with degeneracy pressure at these temperatures and densities, and as a result the torus has a density smaller than the incoming accretion stream.

Because the accretion stream and the helium torus do not move in the same direction, the difference in velocity at the interface is substantial, $U_{1} - U_{2} \sim \sqrt{GM_{1}/r_{1}}(\sqrt{2} - \cos \theta)$, where $\theta$ is the angle between the stream and torus velocities. This shear can lead to the rapid growth of Kelvin-Helmholtz instabilities along the interface between the two regions (Figure \ref{slicefig}, region II). An important factor in determining which perturbations are unstable is the magnitude of the restoring force perpendicular to this interface. The acceleration $g$ provides stability for perturbations with $k < k_{\min}$ \citep{Chandrasekhar:1961p3416}
\begin{equation}
k_{\min} = \frac{g \left(\alpha_{1}-\alpha_{2}\right)}{\alpha_{1}\alpha_{2}\left(U_{1} - U_{2}\right)^{2}},
\end{equation}
where $\alpha_{1} \equiv \rho_{1}/(\rho_{1}+\rho_{2})$ and $\alpha_{2} \equiv \rho_{2}/(\rho_{1}+\rho_{2})$. As long as $g > 0$, perturbations with wavelength $\lambda > \lambda_{\max} = 2\pi/k_{\min}$ will not grow. Within the primary's Roche lobe, $g$ is determined by the combination of the primary's gravity and the Coriolis force
\begin{equation}
g = 2 \Omega |\vec{v}| - \frac{G M_{1}}{\left|\vec{r}_{1}\right|^{2}}\frac{\left|\vec{r}_{1}\times \vec{v}\right|}{\left|\vec{r}_{1}\right|\left|\vec{v}\right|}.
\end{equation}
Because the Coriolis force is always perpendicular to the velocity, its contribution to $g$ is always equal to the magnitude of the force. Only the component of the primary's gravity that is perpendicular to $\vec{v}$ will contribute to $g$. Near {\it L1} where the cumulative gravitational forces are parallel to the velocity, the Coriolis term is dominant and is directed away from the primary, and thus $g$ is positive. As the stream falls toward the surface of the primary, $U_{1} - U_{2}$ continually increases, and the restriction on $k$ is progressively less limiting. And because the Coriolis effect is $\propto |\vec{v}| \propto |\vec{r}_{1}|^{-1/2}$ while the primary's gravity is $\propto |\vec{r}_{1}|^{-2}$, the primary's gravity eventually becomes the dominant contributor to the restoring force. As the primary's gravity is opposite to the Coriolis force, $g$ eventually changes sign, at which point all wave modes become unstable (Figure \ref{slicefig}, region III).

The helium torus, which is both clumpy and has a radial density profile, drives the seed vertical displacements in the accretion stream. The torus is highly turbulent, with the largest bulk motions being driven at scales comparable to the width of the accretion stream ($\sim 10^{8}$ cm) and at speeds that are highly supersonic. Therefore, the power spectrum of the resulting turbulent cascade is given by Burgers turbulence ($E \propto k^{-2}$). Consequently, perturbations with large wavelengths are seeded with large initial amplitudes. Additionally, power is transferred from large $k$ to small $k$ by the tidal stretching of waves as they fall in the primary's gravity. A numerical calculation of the growth rate of the instabilities show that even the largest unstable modes, which have the slowest growth rates, experience a ten-fold increase in amplitude in the time it takes to fall from {\it L1} to the primary's surface. Because the flow is three-dimensional, instabilities larger than the stream size cannot form, and the end result is that the stream consists of knots of material that are comparable in size to the stream itself (Figure \ref{slicefig}).

\begin{figure}[tb]
\centering\includegraphics[width=\linewidth,clip=true]{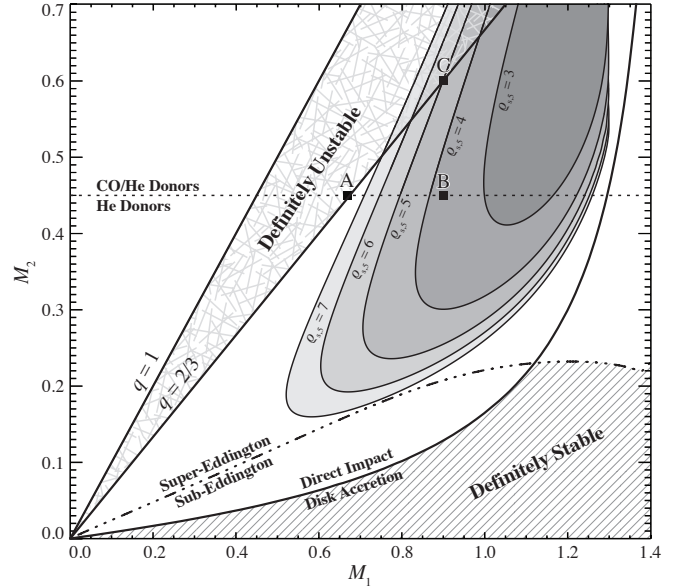}
\caption{The hashed region on the lower right shows systems that are definitely dynamically stable, while the sticked region in the upper left shows systems that are definitely unstable. The thick solid line shows the transition from direct impact to disk accretion, while the dot-dashed curve shows the separation between sub- and super-Eddington accretion in the weak coupling limit \citep{Marsh:2004p1880}. The contours show where $\tau_{\rm dyn} = \tau_{3\alpha}$ for different stream densities $\rho_{\rm s,5} \equiv \rho_{\rm s}/10^{5}$ assuming a torus density of $3 \times 10^{5}$ g cm$^{-3}$. Surface detonations are likely for systems that accrete with $\rho_{\rm s}$ for which $\tau_{\rm dyn} < \tau_{3\alpha}$ and then experience an increase in $\rho_{\rm s}$ until $\tau_{\rm dyn} > \tau_{3\alpha}$.}
\label{stabilityfig}
\end{figure}

\begin{figure*}[tb]
\centering\includegraphics[width=0.475\linewidth,clip=true,angle=-90]{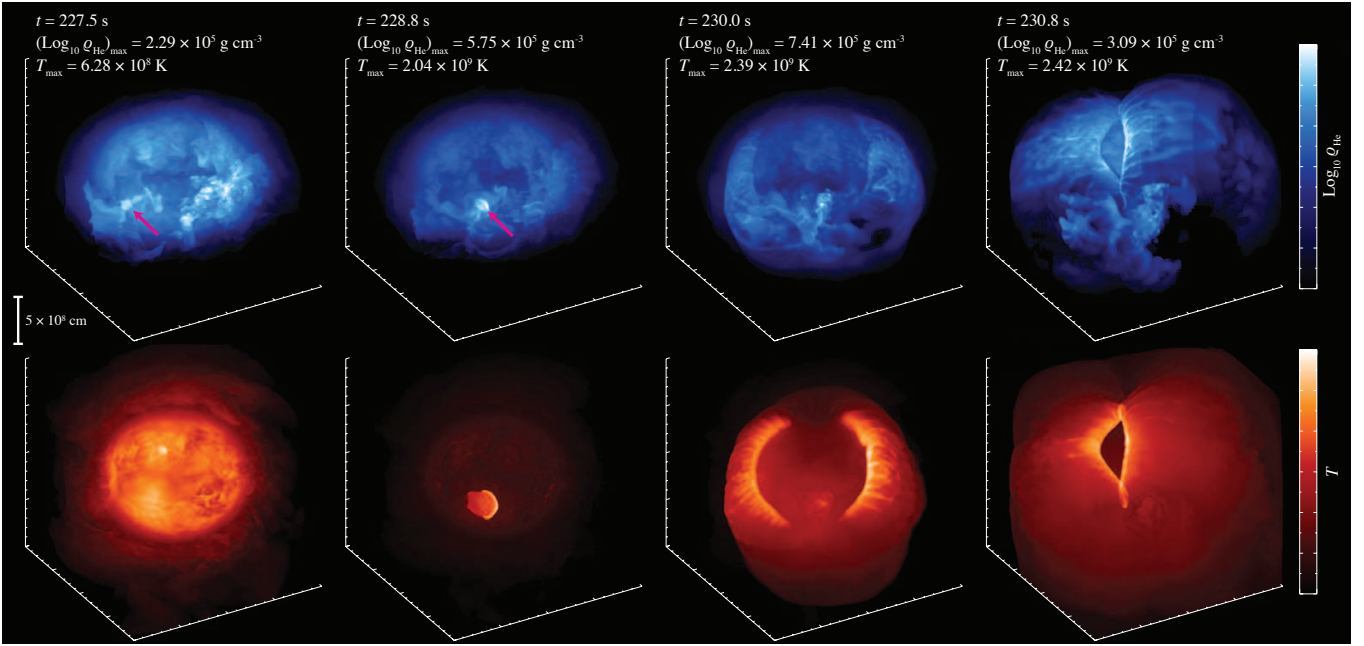}
\caption{Volumetric snapshots showing the temperature $T$ and $\log\rho_{{\rm He}} \equiv \rho X_{{\rm He}}$ during the explosive event on the surface of the primary in run C. A dense knot of helium that formed through Kelvin-Helmholtz instabilities in the accretion stream is marked by a magenta arrow. The evolution of the surface detonation is described in Section \ref{results}.}
\label{volfig}
\end{figure*}

\section{Surface Detonations}

We find that instabilities in the accretion stream are present in all four runs once the helium torus has acquired enough mass to deflect the stream. The material that is compressed against the surface of the primary by the dense knots formed in the stream can be heated to billions of degrees Kelvin (Figure \ref{slicefig}, region IV). At the onset of mass transfer, the density of the stream is too small to heat the underlying material appreciably. This allows $\sim 1-15\%$ of a solar mass of helium to accumulate on the surface of the primary. As the density of the accretion stream increases with the decreasing separation between the WDs, the temperature of the compressed layer can exceed the temperature required to explosively ignite helium via the triple-alpha process. This leads to a gravitationally confined surface detonation that propagates through the helium torus that wraps around the surface of the primary at approximately the Keplerian velocity (Figure \ref{volfig}).

The conditions required for a successful surface detonations are summarized in Figure \ref{stabilityfig}. Only systems that are dynamically unstable will ever have accretion rates capable of igniting surface detonations via stream instabilities, and thus the lower-right region of the figure is excluded. The question of whether the surface detonation will occur in a given binary is tricky to answer because the process is highly stochastic; the impact trajectory, density, and geometry of the knots that initiate the detonations vary considerably over the course of the simulations. Nevertheless, we can estimate when surface detonations will occur by comparing the timescale for triple-$\alpha$ reactions $\tau_{3\alpha}$ \citep{Caughlan:1988p3379} to the dynamical timescale $\tau_{\rm dyn} \simeq R_{1}/v_{\rm esc,1}$ at the surface of the primary.

For a surface detonation to involve a significant amount of mass, the system must accrete material at a rate for which $\tau_{3\alpha} > \tau_{\rm dyn}$, and then transition to a rate where $\tau_{3\alpha} \le \tau_{\rm dyn}$. By solving the equations of motion for a test particle released from {\it L1}, we estimate the two timescales in Figure \ref{stabilityfig} for various stream densities for different primary and secondary masses. The conditions at the He/CO interface are given by $\partial P/\partial r = \partial \phi/\partial r$, where $\phi \simeq G M_{1} / r_{1} - r_{1}\Omega_{\rm tor}^{2}$ and $\Omega_{\rm tor}$ is the angular velocity of the torus, which is related to the component of the accretion stream velocity parallel to the primary's surface. Conversely, the ram pressure $P_{\rm ram} \propto \rho v^{2}$ applied by the accretion stream is dependent on the component of velocity perpendicular to the primary's surface.

The density structure of the torus is approximately fixed as the system evolves towards merger because $\phi$ only increases slightly due to mass accretion. Meanwhile the stream density continually increases, resulting in the contour of $\tau_{3\alpha} = \tau_{\rm dyn}$ moving leftwards in Figure \ref{stabilityfig}. When $\tau_{3\alpha} \lesssim \tau_{\rm dyn}$, the increase in temperature of the torus can no longer be controlled by adiabatic expansion, which results in thermonuclear runaway and a detonation in the helium torus.

\section{Results and Discussion}\label{results}

Surface detonations are present in both of the longer runs with a 0.9 $M_{\odot}$ CO primary, although run B detonates much later in its evolution than run C. The geometry and evolution of the detonations are similar in both runs. A small region of the surface helium is heated to a temperature of $\sim 2 \times 10^{9}$ K, leading to a detonation front that expands outwards from the ignition site along the primary's surface (Figure \ref{volfig}, second column). Because the helium layer is toroidal rather than spherical, the front runs out of fuel as it propagates towards the primary's poles. Consequently, the detonation splits into two fronts that run clockwise and counterclockwise around the equator of the primary (Figure \ref{volfig}, third column). These detonation fronts eventually run into each other along a longitudinal line that is opposite to the original ignition site (Figure \ref{volfig}, fourth column). The highest temperatures are produced in this convergence region.

For run A, the primary's surface gravity is too low to compress the helium layer above the critical temperature necessary for thermonuclear runway. No detonation was observed in run Ba, despite being initialized using a checkpoint from run B just prior to the observed detonation in B. This is because of the stochastic nature of the ignition mechanism --- the particular dense knot of material that led to the surface detonation in B has a different shape in Ba, and did not have a favorable geometry for igniting the helium torus. And because Ba has twice the linear resolution of run B, we could only afford to evolve it for a short period of time.

\begin{figure}[tb]
\centering\includegraphics[width=\linewidth,clip=true]{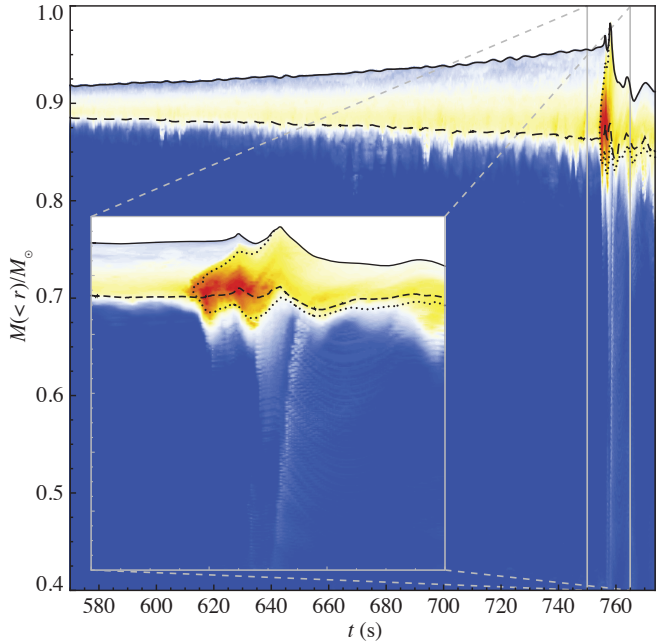}
\caption{Maximum temperature $T$ in a mass shell as a function of interior mass $M\left(< r\right)$ and time $t$ for run B. Each shell lies on a surface of equal gravitational potential. The shading is logarithmically binned, with red corresponding to $T_{\max}$ and blue corresponding to $T_{\min} = T_{\max}/10^{1.5}$. The dashed line shows the region where $X_{\rm He} > 0.02$, while the dotted line shows where $X_{\rm Ne} > 10^{-3}$. For convenience, we include a zoomed view of the surface detonation as an inset.}
\label{radialfig}
\end{figure}

If a helium surface detonation occurs as the result of stream instabilities in a DD system, the resulting transient event could potential resemble a dim Type Ia SNe \citep{Bildsten:2007p2743, Perets:2009p3232}. The results of our simulations shows a large degree of variability which ultimately depends on the geometry of the dense knots when they strike the surface of the primary in the impact zone. However, if the post-detonation temperature is not much larger than $2 \times 10^{9}$ K or the geometry of the detonation fronts are not favorable, it is possible to only synthesize intermediate-mass $\alpha$-elements (Table \ref{outcomes}).

The prospect of igniting the CO core itself is attractive because the typical densities of moderate-mass WDs ($M \sim 1.0 M_{\odot}$) are $\sim 10^{6}-10^{7}$ g cm$^{-3}$, comparable to the densities found in the DDT for Type Ia SNe \citep{Khokhlov:1997p3443}. \cite{Fink:2007p1886} (hereafter FHR) show that when $0.1 M_{\odot}$ of helium is ignited at the He/CO interface of a moderate-mass WD, strong shocks are launched deep into the CO core, which converge at a focusing point. This focusing produces a region where the temperature and density meet the criteria for explosively igniting carbon \citep{Niemeyer:1997p1901, Ropke:2007p3332}, thus leading to a full detonation of the CO core.

The linear resolution of our simulations is substantially coarser than the highest-resolution two-dimensional simulations of FHR, and we are certainly under-resolving the true density and temperature peaks. However, FHR's results appear to be rather optimistic when applied to our model for the following reasons. First, FHR assumes perfect mirror symmetry, which dramatically increases the amount of focusing by forcing the shocks to converge at a single point rather than a locus of points. Second, FHR does not use a nuclear network and makes the simplifying assumption that the entire He envelope is burned to Ni, which naturally overestimates the energy injected into the CO core when compared to the more realistic case where burning is incomplete.

Because of these differences, the conditions for double detonation are still not met even in our run with the most favorable shock geometry and strongest surface detonation (run C). In the focusing region of run C, the temperature and density are $2.5 \times 10^{8}$ K and $2.0 \times 10^{7}$ g cm$^{-3}$, respectively. Conditions for CO core detonation may be more propitious in systems containing a slightly more massive primary, but ultimately the successful detonation of the core depends on the precise distribution of temperature and density within the high-pressure regions of the convergence zone \citep{Seitenzahl:2009p1881}.

The mechanism reported here for the detonation of a sub-Chandrasekhar CO WD in a dynamically unstable binary is not tied to a particular mass scale and therefore allows for considerably more diversity. As outlined above, mass transfer between a pure He WD or a He/CO hybrid and a CO WD before the merger provides a novel pathway to ignite CO WDs. Even if a critical amount of mass is not raised above the conditions required for CO detonation, a peculiar underluminous optical transient should signal the last few orbits of a merging system.

\acknowledgments We have benefited from many useful discussions with C. Fryer, D. Kasen, M. Krumholz, D. Lin, G. Nelemans, and J. Tohline. The
software used in this work was in part developed by the DOE-supported
FLASH center at the University of Chicago. This work is supported by NSF: PHY-0503584, DOE SciDAC: DE-FC02-01ER41176, the Packard Foundation (JG and ER-R) and DFG RO 3399/4-1, 548319 (MD and SR).

\bibliographystyle{apj}
\bibliography{apj-jour,2009a}

\clearpage

\end{document}